# High 3rd Order Optical Kerr Nonlinearity of PdSe$_2$ Di-chalcogenide 2D Films for Nonlinear Photonic Chips

David J. Moss

*Optical Sciences Centre, Swinburne University of Technology, Hawthorn, VIC 3122, Australia*

## ABSTRACT

As a novel layered noble metal dichalcogenide material, palladium diselenide (PdSe$_2$) has attracted wide interest due to its excellent optical and electronic properties. In this work, a strong third-order nonlinear optical response of 2D PdSe$_2$ films is reported. We conduct both open-aperture (OA) and closed-aperture (CA) Z-scan measurements with a femtosecond pulsed laser at 800 nm to investigate the nonlinear absorption and nonlinear refraction, respectively. In the OA experiment, we observe optical limiting behaviour originating from large two photo absorption (TPA) in the PdSe$_2$ film of $\beta$ = 3.26 ×10$^{-8}$ m/W. In the CA experiment, we measure a peak-valley response corresponding to a large and negative Kerr nonlinearity of $n_2$ = -1.33×10$^{-15}$ m$^2$/W – two orders of magnitude larger than bulk silicon. In addition, the variation of $n_2$ as a function of laser intensity is also characterized, with $n_2$ decreasing in magnitude when increasing incident laser intensity, becoming saturated at $n_2$ = -9.96×10$^{-16}$ m$^2$/W at high intensities. Our results show that the extraordinary third-order nonlinear optical properties of PdSe$_2$ have strong potential for high-performance nonlinear photonic devices.

## 1. INTRODUCTION

Two-dimensional (2D) layered materials such as graphene, [1-3] graphene oxide (GO), [4-9] transition metal dichalcogenides (TMDCs), [10-12] and black phosphorus (BP) [13,14] have attracted a great deal of interest, enabling diverse nonlinear photonic devices with vastly superior performance compared to bulk materials. Amongst them, TMDCs (MX$_2$, M = transition metal and X = chalcogen), with bandgaps in the near infrared to the visible region, have opened up promising new avenues for photonic and optoelectronic devices. [2,15,16] For instance, a few mono-layers of MoS$_2$ and WS$_2$ have been used as broadband, fast-recovery saturable absorbers for mode locking in pulsed fiber lasers. [2,15] Nonlinear optical modulators and polarization dependent all-optical switching devices have been realized based on ReSe$_2$ [16] and SnSe. [17] As a new 2D noble metal dichalcogenide in the TMDC family, PdSe$_2$ has recently attracted significant interest. [18-21] Similar to the puckered structure of BP, it has a puckered pentagonal atomic structure − with one Pd atom bonding to four Se atoms and two adjacent Se atoms covalently bonding with each other. Due to this low-symmetry structure, PdSe$_2$ possesses unique in-plane anisotropic optical and electronic properties, [18,19] featuring an in-plane noncentrosymmetric structure, in contrast to its cousin PtSe$_2$. Further, PdSe$_2$ has a layer-dependent bandgap, varying from 0 eV (bulk) to 1.3 eV (monolayer) - a property well suited for photonic and optoelectronic applications – in particular, for wavelength tuneable devices. Moreover, different to BP which degrades rapidly under ambient conditions, PdSe$_2$ is highly air-stable, indicating its robustness and potential for practical applications. The high carrier mobility and anisotropic Raman spectroscopy of 2D PdSe$_2$ layers have been investigated [18,20] as well as highly-sensitive photodetectors from the visible to mid-infrared wavelengths. [22,23] Recently, the optical nonlinear absorption of PdSe$_2$ nanosheets has also been reported in the context of mode-locked laser applications. [24,25] To date, however, its optical Kerr nonlinearity has not been investigated.

Here, we characterize the third-order nonlinear optical properties of PdSe$_2$ multilayer films via Z-scan technique with femtosecond optical pulses at 800 nm. Both OA and CA measurements are performed to investigate the nonlinear absorption and nonlinear refraction of PdSe$_2$. Experimental results show that PdSe$_2$ films exhibit a large and negative (self-defocusing) Kerr nonlinearity ($n_2$) of ∼ -1.33×10$^{-15}$ m$^2$/W, two orders of magnitude larger than bulk silicon. In the OA measurement, we observe a large nonlinear absorption $\beta$ of ∼ 3.26 ×10$^{-8}$ m/W, which originates from TPA in the PdSe$_2$ films. In addition, we investigate the intensity dependence of the nonlinear response of PdSe$_2$, finding that the absolute magnitude of the Kerr nonlinearity $n_2$ initially decreases slightly with incident laser intensity, becoming saturated at higher intensities. These results verify the large third-order nonlinear optical response of PdSe$_2$ as well as its strong potential for high-performance nonlinear photonic devices.

## 2. MATERIAL PREPARATION AND CHARACTERIZATION

The atomic structure of PdSe$_2$ crystals is shown in Figure 1(a). PdSe$_2$ exhibits a unique puckered pentagonal structure, different to other TMDCs like MoS$_2$ and WS$_2$. The Se-Pd-Se layers stack with weak van der Waals interactions to form a layered structure. [18,19] In each monolayer, the pentagonal rings are formed with one Pd atom bonding to four Se atoms and two adjacent Se atoms covalently bonding with each other, which is similar to the puckered structure of BP, and yields both anisotropic and non-centrosymmetric properties of PdSe$_2$. More importantly, unlike the rapid degradation of BP under ambient conditions, PdSe$_2$ has significantly better air-stability. [22,23] Together, these properties of PdSe$_2$ make it promising for high performance photonic and optoelectronic applications.

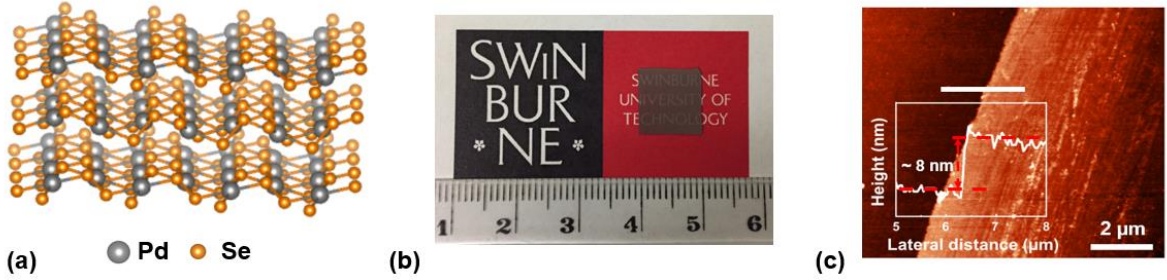

Figure 1. (a) Crystal structure of PdSe$_2$. (b) Photograph of prepared multilayer PdSe$_2$ film. The unit for the numbers on the ruler is centimeter. (c) AFM height profile of the multilayer PdSe$_2$ film.

Here, we investigate large-area multilayer PdSe$_2$ films deposited on transparent sapphire substrates. The PdSe$_2$ films were synthesized via Chemical vapor deposition (CVD). [26] The films were polycrystalline, as is typical for CVD synthesized films, with crystal sizes varying from 10's of nanometres up to 100 nm. Because of the polycrystalline nature of the films, the inversion symmetry breaking properties (i.e., non-centrosymmetric) of the single PdSe$_2$ crystals could not be observed on optical wavelength scales in the macroscopic PdSe$_2$ continuous films studied in this work. Figure 1(b) shows the photography of the prepared PdSe$_2$ film. The morphology image and height profile of the prepared PdSe$_2$ films were characterized by atomic force microscopy (AFM), as illustrated in Figure 1(c). The film thickness was measured to be ~ 8 nm, which corresponds to ~20 layers of PdSe$_2$. [19,26]

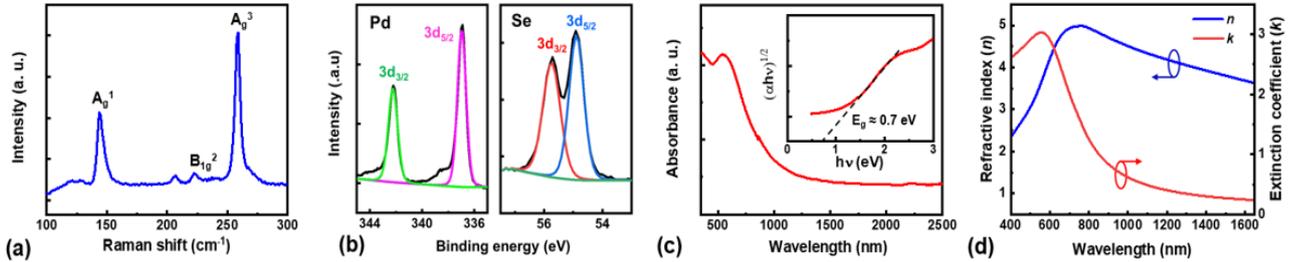

Figure 2. Characterization of the prepared PdSe$_2$ film. (a) Raman spectrum excited via a 514-nm laser. (b) X-ray photoelectron spectroscopy (XPS) spectra. (c) UV-vis absorption spectrum. Inset shows the extracted Tauc plot. (d) Measured refractive index ($n$) and extinction coefficient ($k$).

Raman spectrum of the prepared PdSe$_2$ film excited with a laser at 514 nm is shown in Figure 2(a). Three representative phonon modes can be observed, including the $A_g^1$ (~145.5 cm$^{-1}$) and $B_{1g}^2$ (~222.5 cm$^{-1}$) vibrational modes that correspond to the movement of Se atoms and the $A_g^3$ (~ 258.8 cm$^{-1}$) mode that relates to the relative movements between Pd and Se atoms. [20,26] To further characterize the film quality, X-ray photoelectron spectroscopy (XPS) was employed to measure the binding energy of PdSe$_2$. Figure 2(b) shows the XPS results of Pd 3d and Se 3d core levels for the PdSe$_2$. The peaks of the fit at ~ 342.2 eV and ~ 336.9 eV are attributed to the Pd 3d$_{3/2}$ and Pd 3d$_{5/2}$, respectively, whereas the peaks at ~ 55.7 eV and ~ 54.9 eV correspond to Se 3d$_{3/2}$ and 3d$_{5/2}$, respectively. [20,26] To characterize the linear absorption and optical bandgap, the optical absorption spectrum (from 400 nm to 2500 nm) of the PdSe$_2$ film was measured with ultraviolet-visible (UV-vis) spectrometry, as shown in Figure 2(c). The inset of Figure 2(c) shows the Tauc plot extracted from the linear absorption spectrum, where the optical bandgap of the PdSe$_2$ film is estimated to be ~ 0.7 eV. We also characterize the in-plane (TE-polarized) refractive index ($n$) and extinction coefficient ($k$) of the PdSe$_2$ film via spectral ellipsometry, as depicted in Figure 2(d). The refractive index first increases dramatically with wavelength to reach a peak at ~ 700 nm and then decreases more gradually at longer wavelengths. The extinction coefficient exhibits a significant decrease from 600 nm to 1200 nm, and then the rate of decrease slows down at longer wavelengths. This shows an agreement with the trend of the UV-vis absorption spectrum in Figure 2(c).

## 3. Z-SCAN MEASUREMENTS

To investigate the third-order nonlinear optical properties of $PdSe_2$, we characterized the nonlinear absorption and refraction of the prepared $PdSe_2$ films via the Z-scan technique, [27-29] where a femtosecond pulsed laser with a centre wavelength at ~800 nm and pulse duration of ~ 140 fs was used to excite the samples. A half-wave plate combined with a linear polarizer was employed to control the power of the incident light. A beam expansion system consisting of a 25-mm concave lens and 150-mm convex lens was used to expand the light beam, which was then focused by an objective lens (10 ×, 0.25 NA) to achieve a low beam waist with a focal spot size of ~1.6 μm. The prepared samples were oriented perpendicular to the beam axis and translated along the Z-axis with a linear motorized stage. A high-definition charge-coupled-device (CCD) imaging system was used to align the light beam to the target sample. Two photodetectors (PDs) were employed to detect the transmitted light power for the signal and reference arms.

Figure 3(a) shows the OA Z-scan results for the $PdSe_2$ film at three representative intensities. A typical optical limiting behaviour was observed in the OA curves, with the transmission decreasing as the $PdSe_2$ sample was moved through the focal point. We measured the response of pure sapphire substrate and did not observe any significant nonlinear absorption, indicating that the observed optical limiting response was induced by the $PdSe_2$ film. We also note that the transmittance dip of the OA curve decreased when the incident laser intensity was increased. In principle, the optical limiting behaviour can be induced by several mechanisms such as nonlinear light scattering (NLS), reverse saturable absorption (RSA), two-photon absorption (TPA) and multi-photon absorption (MPA). [30,31] However, apart from the basic condition that the total energy of the photons involved in each process (eg., two photons, for TPA, one photon for SA etc.) must be larger than the bandgap, there is no a-priori reason for any particular process to dominate. For thin $PdSe_2$ film in our case, though, we first exclude the NLS effect since it usually dominates for dispersion or solution samples with laser-induced microbubbles. [30,31] According to the UV-vis spectra, the bandgap of the few-layer $PdSe_2$ film is estimated to be 0.7 eV, which is lower than a single photon energy of the incident laser at 800 nm. Therefore, all the above processes can occur. While SA at low laser intensities and RSA at high laser intensities might be expected for the Z-scan measurements, we did not observe this. This could possibly be because the single photon transition is inefficient under 800-nm femtosecond laser excitation due to the indirect band structure of the few-layer 8-nm-thick $PdSe_2$ films, or possibly parallel band absorption effects. [31] Considering this, RSA is unlikely to dominate the nonlinear absorption in $PdSe_2$ films due to its one photon process. Given the high peak power of the incident femtosecond pulses, TPA is likely to account for the optical limiting behaviour observed in our Z-scan measurements.

To extract the TPA coefficient $β$ of $PdSe_2$, we fit the measured OA results with the well-established theory. [27,28] The TPA coefficient $β$ for the $PdSe_2$ film is shown in Figure 3(b) at different laser intensities. A large $β = 3.26 \times 10^{-8}$ m/W is observed, which is comparable to the reported values of graphene, and higher than that of $WS_2$, highlighting the strong optical limiting effect in $PdSe_2$ film. In addition, the TPA coefficient $β$ is relatively constant with incident laser intensity, reflecting the fact that we are working in an intensity regime where the material properties of the $PdSe_2$ films are not changing much. The slight fluctuation in $β$ with laser intensity may arise from light scattering in the $PdSe_2$ film surface.

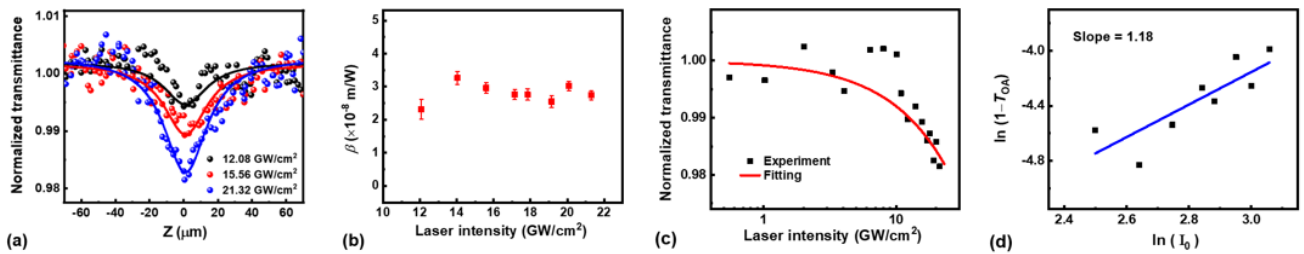

Figure 3. (a) OA Z-scan results of $PdSe_2$ film at different intensities. (b) TPA coefficient $β$ of $PdSe_2$ film versus laser intensity. (c) Normalized transmittance of $PdSe_2$ film at the focal point as a function of laser intensity. (d) The plot of $\ln(1−T_{OA})$ versus $\ln(I_0)$ to determine the order of nonlinearity. The measured and fit results are shown by scatter data points and solid lines, respectively.

To further investigate the nonlinear absorption of the $PdSe_2$ film, we measured the minimum transmittance with the sample at the focal point of the Z scan setup, for different incident laser intensities. Figure 3(c) shows the transmittance of

PdSe$_2$ at the focal point as a function of laser intensity, where the transmittance fluctuates around a relatively constant value at low intensities and then decreases significantly as the laser intensity increased. The experimental data fits the theory well, [31] verifying our assumption of TPA being the dominant process for nonlinear absorption in the PdSe$_2$ film. The order of the observed nonlinear absorption can also be confirmed by examining the relation between ln(1–$T_{OA}$) versus ln($I_0$): [32]

$$\ln(1 - T_{OA}) = k\ln(I_0) + C, \quad (1)$$

where $k$ is the slope showing the order of the nonlinear absorption and $C$ is a constant. For pure TPA, the slope is equal to 1. [32] We obtain a slope of 1.18 (Figure 3(d)), suggesting the observed nonlinear absorption is mainly attributed to TPA in the PdSe$_2$ film.

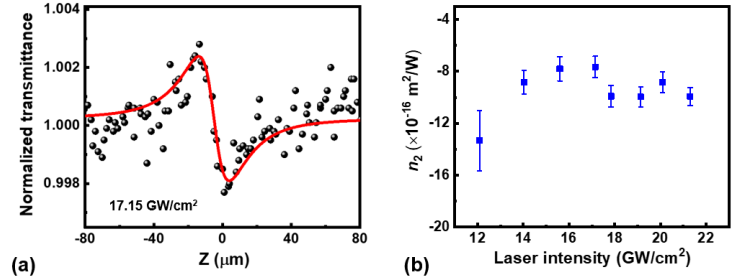

Figure 4. (a) Representative CA Z-scan result of PdSe$_2$ film at intensity of 17.15 GW/cm$^2$. (b) Kerr coefficient $n_2$ of PdSe$_2$ film versus laser intensity.

We also performed CA Z-scan measurements to investigate the Kerr nonlinearity ($n_2$) of the PdSe$_2$ films. The values of $n_2$ for the PdSe$_2$ film at different laser intensities were extracted by fitting the measured CA results. Figure 4(a) shows a representative CA result for PdSe$_2$ at a laser intensity of 17.15 GW/cm$^2$. The transmittance of the sample exhibited a transition from peak to valley when the sample passed through the focal plane. Such a peak-valley CA behaviour corresponds to a negative Kerr coefficient $n_2$ and indicates an optical self-defocusing effect in the PdSe$_2$ film. The noise in the CA data is mainly induced by the light scattering in the PdSe$_2$ film surface. By improving the film uniformity, such noise can be further reduced. As discussed above, TPA results in the transfer of electrons from valence band to conduction band, increasing the free carrier density in the film. Therefore, the observed negative Kerr nonlinearity potentially originates from the TPA-induced free carrier nonlinear refraction and interband blocking. [33,34]

Figure 4(b) shows the measured Kerr coefficient $n_2$ of PdSe$_2$ versus laser intensity, showing a large $n_2$ of −1.33×10$^{-15}$ m$^2$/W. Table 1 compares the $\beta$ and $n_2$ of PdSe$_2$ with other 2D layered materials. As can be seen, the value of $n_2$ for PdSe$_2$ is lower than those of graphene and GO, but still more than two orders of magnitude higher than bulk silicon. [35, 36] Such a high $n_2$ suggests that PdSe$_2$ is an extremely promising material for self-defocusing based nonlinear photonic applications. For example, a negative Kerr nonlinearity can be used to self-compress ultrashort pulses in the presence of positive group-velocity dispersion. [37] Another application of a negative Kerr nonlinearity is mode locking of lasers using the Kerr mode-locking technique [35, 38] as well as the possibility of achieving net parametric modulational instability gain under normal dispersion conditions. [35, 39]

In addition, as shown in Figure 4(b), the absolute value of $n_2$ initially decreases with laser intensity and then saturates at higher intensities. In theory, the optical nonlinear refraction originates mainly from the free-carrier and bound-electron nonlinearities. [33, 40-44] We assume that the two mechanisms co-exist in the PdSe$_2$ film. It has been shown that, near the half-bandgap the two-photon resonance typically yields a positive $n_2$. However, at higher photon energies, the bound-electron contribution to the $n_2$ nonlinearity becomes negative, while the free-carrier nonlinearity is usually also negative. [33,44] We therefore infer that either, or both, processes contribute to the nonlinearity since we observed a negative Kerr nonlinearity for the PdSe$_2$ film. This is further complicated by the fact that PdSe$_2$ is an indirect bandgap material. The Kerr nonlinearity is dominated by direct transitions at all energies, whereas the nonlinear absorption is dominated by indirect transitions in energy regions where the direct transitions are not allowed (eg., below 1/2 the direct bandgap for TPA). [45]

The refractive index change in the PdSe$_2$ film can be expressed by $\Delta n = n_2^* I_0 + \sigma_r N$, where $n_2^*$ is the nonlinear refraction related to bonding electrons, $\sigma_r$ is the free carrier refractive coefficient and $N$ is the charge carrier density. [33] Therefore, the effective $n_2 = \Delta n/I_0 = n_2^* + \sigma_r N/I_0$, is an intensity dependent parameter, which can explain the $n_2$ variation as a function of laser intensity observed in our measurements.

Finally, nonlinear devices rely on parametric gain and FWM in MRRs which depend on many factors in terms of material properties, including the third-order nonlinearity, the linear and nonlinear loss, dispersion, etc. The extremely promising nonlinear optical properties of layered GO films will yield many new device properties that are difficult to achieve for typical integrated photonic devices. We believe this could enable one to tailor the device performance for many applications to microcomb devices, quantum optics and nonlinear optical photonic chips in general [58-164].

Table 1. Comparison of $\beta$ and $n_2$ for various 2D layered materials

| Material | Laser parameter | Thickness | $\beta$ (m/W) | $n_2$ (m$^2$/W) | $n_2$ (×$n_2$ of Si[1]) | Ref. |
|---|---|---|---|---|---|---|
| Graphene | 1150 nm, 100 fs | 5-7 layers | $3.8 \times 10^{-8}$ | $-5.5 \times 10^{-14}$ | $-1.22 \times 10^4$ | [42] |
| GO | 800 nm, 100 fs | ~2 μm | $4 \times 10^{-7}$ | $1.25 \times 10^{-13}$ | $2.75 \times 10^4$ | [29] |
| MoS$_2$ | 1064 nm, 25 ps | ~25 μm | $(-3.8 \pm 0.59) \times 10^{-11}$ | $(1.88 \pm 0.48) \times 10^{-16}$ | 41.32 | [11] |
| WS$_2$ | 1040 nm, 340 fs | ~57.9 nm | $(1.81 \pm 0.08) \times 10^{-8}$ | $(-3.36 \pm 0.27) \times 10^{-16}$ | -73.85 | [12] |
| BP | 1030 nm, 140 fs | ~15 nm | $5.845 \times 10^{-6}$ | $-1.635 \times 10^{-12}$ | $-3.59 \times 10^5$ | [14] |
| PtSe$_2$ | 800 nm, 150 fs | ~4.6 nm | $-8.80 \times 10^{-8}$ | – | – | [43] |
| PtSe$_2$ | 1030 nm, 340 fs | 17 layers | – | $(-3.76 \pm 0.46) \times 10^{-15}$ | $-8.26 \times 10^2$ | [32] |
| PdSe$_2$ | 800 nm, 140 fs | ~8 nm | $(3.26 \pm 0.19) \times 10^{-8}$ | $(-1.33 \pm 0.23) \times 10^{-15}$ | $-2.92 \times 10^2$ | This work |

[1] $n_2$ for silicon = $4.55 \times 10^{-18}$ m$^2$/W (ref. [36])

## 4. CONCLUSION

In summary, we report a large third-order nonlinear optical response of PdSe$_2$ films measured with the Z-scan technique. Experimental results show that PdSe$_2$ has a strong TPA response with a large $\beta$ of ~ $3.26 \times 10^{-8}$ m/W. The Kerr nonlinearity ($n_2$) of PdSe$_2$ is also investigated. We find that $n_2$ is negative, and with an absolute magnitude that is more than two orders of magnitude larger than bulk silicon. Furthermore, we characterize the variation in $n_2$ of PdSe$_2$ with laser intensity, finding that $n_2$ initially increases (decreasing in absolute magnitude) with incident laser intensity and then saturates at higher intensities. Our results verify PdSe$_2$ as a promising 2D material with prominent nonlinear optical properties similar to graphene oxide [46-49], and potentially even for mid IR applications on platforms such as Si-Ge [50-57].

## ACKNOWLEDGEMENT

This work was supported by the Australian Research Council Discovery Projects Program (No. DP150102972 and DP190103186), and the Industrial Transformation Training Centres scheme (Grant No. IC180100005). We acknowledge Swinburne Nano Lab and Micro Nano Research Facility (MNRF) of RMIT University for the support in material characterization as well as Shenzhen Sixcarbon Technology for the PdSe$_2$ film fabrication. We thank Dr. Yunyi Yang and Dr. Tania Moein for technical support, Dr. Deming Zhu for assisting in XPS characterization and Dr. Chenglong Xu for assisting in optical characterization.

Competing interests: The authors declare no competing interests.

## REFERENCES

[1] Lim, G.-K.; Chen, Z.-L.; Clark, J.; Goh, R. G. S.; Ng, W.-H.; Tan, H.-W.; Friend, R.H.; Ho, P. K. H.; Chua, L.-L. Giant broadband nonlinear optical absorption response in dispersed graphene single sheets. *Nat. Photonics* 2011, *5*, 554-560.
[2] Pan, T.; Qiu, C.; Wu, J.; Jiang, X.; Liu, B.; Yang, Y.; Zhou, H.; Soref, R.; Su, Y. Analysis of an electro-optic modulator based on a graphene-silicon hybrid 1D photonic crystal nanobeam cavity. *Opt. Express* 2015, *23*, 23357-233364.
[3] Wu, J., Jia, L., Zhang, N., Qu, Y., Jia, B. and Moss, D. J. Graphene Oxide for Integrated Photonics and Flat Optics. *Adv. Mater.* **2020**, 32, 1-29, 2006415.
[4] Yang, Y.; Wu, J.; Xu, X.; Liang, Y.; Chu, S. T.; Little, B. E.; Morandotti, R.; Jia, B.; Moss, D. J. Invited Article: Enhanced four-wave mixing in waveguides integrated with graphene oxide. *APL Photonics* 2018, 3, 120803.
[5] Yang, Y.; Lin, H.; Zhang, B. Y.; Zhang, Y.; Zheng, X.; Yu, A.; Hong, M.; Jia, B. Graphene-based multilayered metamaterials with phototunable architecture for on-chip photonic devices. *ACS Photonics* 2019, 6, 1033-1040.
[6] Wu, J.; Yang, Y.; Qu, Y.; Xu, X.; Liang, Y.; Chu, S. T.; Little, B. E.; Morandotti, R.; Jia, B.; Moss, D. J. Graphene oxide waveguide and micro-ring resonator polarizers. *Laser Photonics Rev.* 2019, 13, 1900056.
[7] Wu, J.; Yang, Y.; Qu, Y.; Jia, L.; Zhang, Y.; Xu, X.; Chu, S. T.; Little, B. E.; Morandotti, R.; Jia, B.; Moss, D. J. 2D Layered graphene oxide films integrated with micro-ring resonators for enhanced nonlinear optics. *Small* 2020, 16, 1906563.
[8] Zhang, Y.; Wu, J.; Yang, Y.; Qu, Y.; Jia, L.; Moein, T.; Jia, B.; Moss, D. J. Enhanced Kerr Nonlinearity and Nonlinear Figure of Merit in Silicon Nanowires Integrated with 2D Graphene Oxide Films. *ACS Appl. Mater. Interfaces* 2020, 12, 33094-33103.
[9] Qu, Y., Wu, J., Yang, Y., Zhang, N., Liang, Y., Dirani, H. E., Crochemore, R., Demongodin, P., Sciancalepore, C., Grillet, C., Monat, C., Jia, B., David J. Moss. Enhanced Four-Wave Mixing in Silicon Nitride Waveguides Integrated with 2D Layered Graphene Oxide Films. *Adv. Opt. Mater.* 2020, 8, 2001048.

[10] Chen, H.; Corboliou, V.; Solntsev, A. S.; Choi, D. Y.; Vincenti, M. A.; de Ceglia, D.; de Angelis, C.; Lu, Y.; Neshev, D. N. Enhanced second-harmonic generation from two-dimensional MoSe2 on a silicon waveguide. *Light: Science & Applications* 2017, *6*, e17060.
[11] Bikorimana, S.; Lama, P.; Walser, A.; Dorsinville, R.; Anghel, S.; Mitioglu, A.; Micu, A.; Kulyuk, L. Nonlinear optical response in two-dimensional transition metal dichalcogenide multilayer: WS2, WSe2, MoS2 and Mo0.5W0.5S2. *Opt. Express* 2016, 24, 20685-20695.
[12] Dong, N.; Li, Y.; Zhang, S.; McEvoy, N.; Zhang, X.; Cui, Y.; Zhang, L.; Duesberg, G. S.; Wang, J. Dispersion of nonlinear refractive index in layered WS2 and WSe2 semiconductor films induced by two-photon absorption. *Opt. Lett.* 2016, 41, 3936-3939.
[13] Zheng, X.; Chen, R.; Shi, G.; Zhang, J.; Xu, Z.; Cheng, X.; Jiang, T. Characterization of nonlinear properties of black phosphorus nanoplatelets with femtosecond pulsed Z-scan measurements. *Opt. Lett.* 2015, *40*, 3480.
[14] Yang, T.; Abdelwahab, I.; Lin, H.; Bao, Y.; Tan, S. J. R.; Fraser, S.; Loh, K. P.; Jia, B. Anisotropic third-order nonlinearity in pristine and lithium hydride intercalated black phosphorus *ACS Photonics* 2018, *5*, 4969-4977.
[15] Chen, B.; Zhang, X.; Wu, K.; Wang, H.; Wang, J.; Chen, J. Q-switched fiber laser based on transition metal dichalcogenides MoS2, MoSe2, WS2, and WSe2. *Opt. Express* **2015**, *23*, 26723-26737.
[16] Du, L.; Jiang, G.; Miao, L.; Huang, B.; Yi, J.; Zhao, C.; Wen, S. Few-layer rhenium diselenide: an ambient-stable nonlinear optical modulator. *Opt. Mater. Express* 2018, 8, 926-935.
[17] Zhang, C.; Ouyang, H.; Miao, R.; Sui, Y.; Hao, H.; Tang, Y.; You, J.; Zheng, X.; Xu, Z.; Cheng, X.; Jiang, T. Anisotropic nonlinear optical properties of a SnSe flake and a novel perspective for the application of all-optical switching. *Adv. Opt. Mater.* 2019, *7*, 1900631.
[18] Oyedele, A. D.; Yang, S.; Liang, L.; Puretzky, A. A.; Wang, K.; Zhang, J.; Yu, P.; Pudasaini, P. R.; Ghosh, A. W.; Liu, Z.; Rouleau, C. M.; Sumpter, B. G.; Chisholm, M. F.; Zhou, W.; Rack, P. D.; Geohegan, D. B.; Xiao, K. PdSe2: Pentagonal two-dimensional layers with high air stability for electronics. *J. Am. Chem. Soc.* 2017, *139*, 14090-14097.
[19] Zhang, G.; Amani, M.; Chaturvedi, A.; Tan, C.; Bullock, J.; Song, X.; Kim, H.; Lien, D.-H.; Scott, M. C.; Zhang, H.; Javey, A. Optical and electrical properties of two-dimensional palladium diselenide. *Appl. Phys. Lett.* 2019, *114*, 253102.
[20] Jiang, S.; Xie, C.; Gu, Y.; Zhang, Q.; Wu, X.; Sun, Y.; Li, W.; Shi, Y.; Zhao, L.; Pan, S.; Yang, P.; Huan, Y.; Xie, D.; Zhang, Q.; Liu, X.; Zou, X.; Gu, L.; Zhang, Y. Anisotropic growth and scanning tunneling microscopy identification of ultrathin even layered PdSe2 ribbons. *Small* 2019, *15*, 1902769.
[21] Jia, L., Wu, J., Yang, T., Jia, B., Moss, D. J. Large Third-Order Optical Kerr Nonlinearity in Nanometer-Thick PdSe2 2D Dichalcogenide Films: Implications for Nonlinear Photonic Devices. *ACS Appl. Nano Mater.* 2020, 3, 6876-6883.
[22] Liang, Q.; Wang, Q.; Zhang, Q.; Wei, J.; Lim, S. X.; Zhu, R.; Hu, J.; Wei, W.; Lee, C.; Sow, C.; Zhang, W.; Wee, A. T. S. High-Performance, room temperature, ultra-broadband photodetectors based on air-stable PdSe2. *Adv. Mater.* 2019, *31*, 1807609.
[23] Zeng, L.-H.; Wu, D.; Lin, S.-H.; Xie, C.; Yuan, H.-Y.; Lu, W.; Lau, S. P.; Chai, Y.; Luo, L.-B.; Li, Z.-J.; Tsang, Y. H. Controlled synthesis of 2D PdSe2 for sensitive photodetector applications. *Adv. Funct. Mater.* 2019, *29*, 1806878.
[24] Ma, Y.; Zhang, S.; Ding, S.; Liu, X.; Yu, X.; Peng, F.; Zhang, Q. Passively Q-switched Nd: GdLaNbO4 laser based on 2D PdSe2 nanosheet. *Optics and Laser Technology*, 2019, *124*, 105959.
[25] Zhang, H.; Ma, P.; Zhu, M.; Zhang, W.; Wang, G.; Fu, S. Palladium selenide as a broadband saturable absorber for ultra-fast photonics. *Nanophotonics* 2020, https://doi.org/10.1515/nanoph-2020-0116.
[26] Xu, H.; Zhang, H.; Liu, Y.; Zhang, S.; Sun, Y.; Guo, Z.; Sheng, Y.; Wang, X.; Luo, C.; Wu, X.; Wang, J.; Hu, W.; Xu, Z.; Sun, Q.; Zhou, P.; Shi, J.; Sun, Z.; Zhang, D. W.; Bao, W. Controlled doping of wafer-scale PtSe2 films for device application. *Adv. Funct. Mater*. 2019, *29*, 1805614.
[27] Jia, L.; Cui, D.; Wu, J.; Feng, H.; Yang, Y.; Yang, T.; Qu, Y.; Du, Y.; Hao, W.; Jia, B.; Moss, D. J. Highly nonlinear BiOBr nanoflakes for hybrid integrated photonics. *APL Photonics* 2019, *4*, 090802.
[28] Sheik-Bahae, M.; Said, A. A.; Wei, T.-H.; Hagan, D. J.; Stryland, E. W. V. Sensitive measurement of optical nonlinearities using a single beam. *IEEE J. Quantum Electron*. 1990, *26*, 760-769.
[29] Zheng, X.; Jia, B.; Chen, X.; Gu, M. In situ third-order nonlinear responses during laser reduction of graphene oxide thin films towards on-chip nonlinear photonic devices. *Adv. Mater.* 2014, *26*, 2699-2703.
[30] Chantharasupawong, P.; Philip, R.; Narayanan, N. T.; Sudeep, P. M.; Mathkar, A.; Ajayan, P. M.; Thomas, J. Optical power limiting in fluorinated graphene oxide: an insight into the nonlinear optical properties. *J. Phys. Chem. C* 2012, *116*, 25955-25961.
[31] Wang, G.; Bennett, D.; Zhang, C.; Ó Coileáin, C.; Liang, M.; McEvoy, N.; Wang, J. J.; Wang, J.; Wang, K.; Nicolosi, V.; Blau, W. J. Two‐photon absorption in monolayer MXenes. *Adv. Opt. Mater*. 2020, *8*, 1902021.
[32] Wang, L.; Zhang, S.; McEvoy, N.; Sun, Y.; Huang, J.; Xie, Y.; Dong, N.; Zhang, X.; Kislyakov, I. M.; Nunzi, J. M.; Zhang, L.; Wang, J. Nonlinear optical signatures of the transition from semiconductor to semimetal in PtSe2. *Laser Photonics Rev*. 2019, *13*, 1900052.
[33] Hagan, D. J.; Van Stryland, E. W.; Soileau, M. J.; Wu, Y. Y. Self-protecting semiconductor optical limiters. *Opt. Lett*., 1988, *13*, 315-317.
[34] Said, A. A.; Sheik-Bahae, M.; Hagan, D. J.; Wei, T. H.; Wang, J.; Young, J.; Van Stryland, E. W. Determination of bound-electronic and free-carrier nonlinearities in ZnSe, GaAs, CdTe, and ZnTe. *J. Opt. Soc. Am. B* 1992, *9*, 405-414.
[35] Pasquazi, A.; Peccianti, M.; Razzari, L.; Moss, D. J.; Coen, S.; Erkintalo, M.; Chembo, Y. K.; Hansson, T.; Wabnitz, S.; Del'Haye, P.; Xue, X.; Weiner, A. M.; Morandotti, R. Micro-Combs: A Novel Generation of Optical Sources. *Physics Reports* 2018, *729*, 1-81.

[36] Moss, D. J.; Morandotti, R.; Gaeta, A. L.; Lipson, M. New CMOS-compatible platforms based on silicon nitride and Hydex for nonlinear optics. *Nat. Photonics* 2013, *7*, 597-607.
[37] DeSalvo, R.; Hagan, D. J.; Sheik-Bahae, M.; Stegeman, G.; Van Stryland, E. W.; Vanherzeele, H. Self-focusing and self-defocusing by cascaded second-order effects in KTP. *Opt. Lett*. 1992, *17*, 28-30.
[38] Kriso, C.; Kress, S.; Munshi, T.; Grossmann, M.; Bek, R.; Jetter, M.; Michler, P.; Stolz, W.; Koch, M.; Rahimi-Iman, A. Microcavity-enhanced Kerr nonlinearity in a vertical-external-cavity surface-emitting laser. *Opt. Express* 2019, *27*, 11914-11929.
[39] Liu, Y.; Xuan, Y.; Xue, X.; Wang, P.-H.; Chen, S.; Metcalf, A. J.; Wang, J.; Leaird, D. E.; Qi, M.; Weiner, A. M. Investigation of mode coupling in normal-dispersion silicon nitride microresonators for Kerr frequency comb generation. *Optica* 2014, *1*, 137-144.
[40] Lu, S.; Zhao, C.; Zou, Y.; Chen, S.; Chen, Y.; Li, Y.; Zhang, H.; Wen, S.; Tang, D. Third order nonlinear optical property of $Bi_2Se_3$. *Opt. Express* 2013, *21*, 2072-2082.
[41] Sheik-Bahae, M.; Hagan, D. J.; Van Stryland, E. W. Dispersion of bound electronic nonlinear refraction in solids. *IEEE J. Quantum Electron.* 1991, *27*, 1296-1309.
[42] Demetriou, G.; Bookey, H. T.; Biancalana, F.; Abraham, E.; Wang, Y.; Ji, W.; Kar, A. K. Nonlinear optical properties of multilayer graphene in the infrared. *Opt. Express* 2016, *24*, 13033-13043.
[43] Wang, G.; Wang, K.; McEvoy, N.; Bai, Z.; Cullen, C. P.; Murphy, C. N.; McManus, J. B.; Magan, J. J.; Smith, C. M.; Duesberg, G. S.; Kaminer, I.; Wang, J.; Blau, W. J. Ultrafast carrier dynamics and bandgap renormalization in layered $PtSe_2$. *Small* 2019, *15*, 1902728.
[44] Huang, J.; Dong, N.; McEvoy, N.; Wang, L.; Coileain, C. O.; Wang, H.; Cullen, C. P.; Chen, C.; Zhang, S.; Zhang, L.; Wang, J. Surface-State Assisted Carrier Recombination and Optical Nonlinearities in Bulk to 2D Nonlayered PtS. *ACS Nano* 2019, *13*, 13390.
[45] Moss, D. J.; Ghahramani, E.; Sipe, J. E.; van Driel, H. M. Band structure calculation of dispersion and anisotropy in $\chi^{(3)}$ for third harmonic generation in Si, Ge, and GaAs. *Phys. Rev. B* 1990, *41*, 1542-1560.
[46] Yang Qu, Jiayang Wu, Yuning Zhang, Yao Liang, Baohua Jia, and David J. Moss, "Analysis of four-wave mixing in silicon nitride waveguides integrated with 2D layered graphene oxide films", Journal of Lightwave Technology **39** Early Access (2021). DOI: 10.1109/JLT.2021.3059721.
[47] Yuning Zhang, Jiayang Wu, Yang Qu, Linnan Jia, Baohua Jia, and David J. Moss, "Analysis of self-phase modulation in silicon-on-insulator nanowires integrated with 2D layered graphene oxide films, Journal of Lightwave Technology (2021).
[48] Moss, David. "Design of silicon waveguides integrated with 2D graphene oxide films for Kerr nonlinear optics". TechRxiv. Preprint. (2020), https://doi.org/10.36227/techrxiv.13203278.v1
[49] Zhang, Y.; Wu, J.; Qu, Y.; Jia, L.; Jia, B.; Moss, D., "Design of Silicon Waveguides Integrated with 2D Graphene Oxide Films for Kerr Nonlinear Optics", Preprints (2020), 2020110279.
[50] Alberto Della Torre, Milan Sinobad, Remi Armand, Barry Luther-Davies, Pan Ma, Stephen Madden, Arnan Mitchel, David J. Moss, Jean-Michel Hartmann, Vincent Reboud, Jean-Marc Fedeli, Christelle Monat, Christian Grillet, "Mid-infrared supercontinuum generation in a low-loss germanium-on-silicon waveguide", APL Photonics **6**, 016102 (2021); doi: 10.1063/5.0033070.
[51] Milan Sinobad, Alberto Della Torre, Remi Armand, Barry Luther-Davies, Pan Ma, Stephen Madden, Arnan Mitchell, David J. Moss, Jean-Michel Hartmann, Jean-Marc Fedeli, Christelle Monat, and Christian Grillet,"Mid-infrared supercontinuum generation in silicon-germanium all-normal dispersion waveguides", Optics Letters **45** (18), 5008-5011 (2020). DOI: 10.1364/OL.402159.
[52] Milan Sinobad, Alberto Della Torre, Barry Luther-Davis, Pan Ma, Stephen Madden, Arnan Mitchell, David J. Moss, Jean-Michel Hartmann, Jean-Marc Fédéli, Christelle Monat and Christian Grillet, "High coherence at f and 2f of a mid-infrared supercontinuum in a silicon germanium waveguide", IEEE Journal of Selected Topics in Quantum Electronics (JSTQE) **26** (2) (March/April) 8201008 (2020). DOI:10.1109/JSTQE.2019.2943358.
[53] Milan Sinobad, Alberto Della Torre, Barry Luther-Davis, Pan Ma, Stephen Madden, Sukanta Debbarma, Khu Vu, David J. Moss, Arnan Mitchell, Regis Orobtchouk, Jean-Marc Fedeli, Christelle Monat, and Christian Grillet, "Dispersion trimming for mid-infrared supercontinuum generation in a hybrid chalcogenide/silicon-germanium waveguide", Journal of the Optical Society of America B (JOSA B) **36**, (2) A98-A104 (2019). DOI: 10.1364/JOSAB.36.000A98.
[54] Milan Sinobad, Christelle Monat, Barry Luther-Davies, Pan Ma, Stephen Madden, David J. Moss, Arnan Mitchell, David Allioux, Regis Orobtchouk, Salim Boutami, Jean-Michel Hartmann, Jean-Marc Fedeli, Christian Grillet, "High brightness mid-infrared octave spanning supercontinuum generation to 8.5μm in chip-based Si-Ge waveguides", Optica **5** (4) 360-366 (2018). DOI:10.1364/OPTICA.5.000360.
[55] L. Carletti, P. Ma, B. Luther-Davies, D. Hudson, C. Monat, S. Madden, M. Brun, S. Ortiz, S. Nicoletti, D. J. Moss, and C. Grillet, "Nonlinear optical properties of SiGe waveguides in the mid-infrared", Optics Express **23**, (7) pp. 8261–8271 (2015).
[56] A. Frigg, A. Boes, G. Ren, T.G. Nguyen, D.Y. Choi, S. Gees, D. Moss, A. Mitchell, "Optical frequency comb generation with low temperature reactive sputtered silicon nitride waveguides", APL Photonics, vol.5, Issue 1, 011302 (2020).
[57] T. Moein et al., "Optically-Thin Broadband Graphene-Membrane Photodetector", Nanomaterials, vol. 10, issue 3, 407 (2020).
[58] L. Razzari, D. Duchesne, M. Ferrera, R. Morandotti, S. Chu, B. E. Little, and D. J. Moss, "CMOS-compatible integrated optical hyper-parametric oscillator," *Nat. Photonics,* vol. 4, no. 1, pp. 41-45, Jan. 2010.
[59] D.J. Moss, R. Morandotti, A. L. Gaeta, and M. Lipson, "New CMOS-compatible platforms based on silicon nitride and Hydex for nonlinear optics," *Nat. Photonics,* vol. 7, no. 8, pp. 597-607, Aug. 2013.
[60] Pasquazi, M. Peccianti, L. Razzari, D. J. Moss, S. Coen, M. Erkintalo, Y. K. Chembo, T. Hansson, S. Wabnitz, P. Del Haye, X. X. Xue, A. M. Weiner, and R. Morandotti, "Micro-Combs: A Novel Generation of Optical Sources", *Physics Reports,* vol. 729, pp. 1-81, Jan. 2018.

[61] M. Ferrera, L. Razzari, D. Duchesne, R. Morandotti, Z. Yang, M. Liscidini, J. E. Sipe, S. Chu, B. E. Little, and D. J. Moss, "Low-power continuous-wave nonlinear optics in doped silica glass integrated waveguide structures," *Nat. Photonics,* vol. 2, no. 12, pp. 737-740, 2008.

[62] M. Kues, C. Reimer, P. Roztocki, L. R. Cortes, S. Sciara, B. Wetzel, Y. B. Zhang, A. Cino, S. T. Chu, B. E. Little, D. J. Moss, L. Caspani, J. Azana, and R. Morandotti, "On-chip generation of high-dimensional entangled quantum states and their coherent control," *Nature,* vol. 546, no. 7660, pp. 622-626, Jun. 2017.

[63] L. Caspani, C. Xiong, B. Eggleton, D. Bajoni, M. Liscidini, M. Galli, R. Morandotti, David J. Moss, "On-chip sources of quantum correlated and entangled photons", *Nature*: Light Science and Applications, vol. 6, e17100 (2017); doi: 10.1038/lsa.2017.100.

[64] M. Kues, C. Reimer, B. Wetzel, P. Roztocki, B. E. Little, S. T. Chu, T. Hansson, E. A. Viktorov, D. J. Moss, and R. Morandotti, "Passively mode-locked laser with an ultra-narrow spectral width," *Nat. Photonics,* vol. 11, no. 9, pp. 608-608, Sep. 2017.

[65] C. Reimer, M. Kues, P. Roztocki, B. Wetzel, F. Grazioso, B. E. Little, S. T. Chu, T. Johnston, Y. Bromberg, L. Caspani, D. J. Moss, and R. Morandotti, "Generation of multiphoton entangled quantum states by means of integrated frequency combs," *Science,* vol. 351, no. 6278, pp. 1176-1180, Mar. 2016.

[66] M. Ferrera, Y. Park, L. Razzari, B. E. Little, S. T. Chu, R. Morandotti, D. J. Moss, and J. Azana, "On-chip CMOS-compatible all-optical integrator," *Nat. Commun.,* vol. 1, Article 29. 1 Jun. 2010. doi:10.1038/ncomms1028

[67] F. Da Ros, E. P. da Silva, D. Zibar, S. T. Chu, B. E. Little, R. Morandotti, M. Galili, D. J. Moss, and L. K. Oxenlowe, "Wavelength conversion of QAM signals in a low loss CMOS compatible spiral waveguide," *APL Photonics*, vol. 2, no. 4, 046105. Apr. 2017.

[68] Pasquazi, L. Caspani, M. Peccianti, M. Clerici, M. Ferrera, L. Razzari, D. Duchesne, B. E. Little, S. T. Chu, D. J. Moss, and R. Morandotti, "Self-locked optical parametric oscillation in a CMOS compatible microring resonator: a route to robust optical frequency comb generation on a chip," *Opt. Express,* vol. 21, no. 11, pp. 13333-13341, Jun. 2013.

[69] Reimer, M. Kues, L. Caspani, B. Wetzel, P. Roztocki, M. Clerici, Y. Jestin, M. Ferrera, M. Peccianti, A. Pasquazi, B. E. Little, S. T. Chu, D. J. Moss, and R. Morandotti, "Cross-polarized photon-pair generation and bi-chromatically pumped optical parametric oscillation on a chip," *Nat. Commun.,* vol. 6, Article 8236. Sep. 2015.

[70] J. Wu, X. Xu, T. G. Nguyen, S. T. Chu, B. E. Little, R. Morandotti, A. Mitchell, and D. J. Moss, "RF Photonics: An Optical Microcombs' Perspective," *IEEE Journal of Selected Topics in Quantum Electronics,* vol. 24, no. 4, pp. 1-20. 2018.

[71] X. Xu, M. Tan, J. Wu, R. Morandotti, A. Mitchell, and D. J. Moss, "Microcomb-based photonic RF signal processing," *IEEE Photonics Technology Letters,* vol. 31, no. 23, pp. 1854-1857. 2019.

[72] X. Xu, J. Wu, T. G. Nguyen, T. Moein, S. T. Chu, B. E. Little, R. Morandotti, A. Mitchell, and D. J. Moss, "Photonic microwave true time delays for phased array antennas using a 49 GHz FSR integrated optical micro-comb source [Invited]," *Photonics Research,* vol. 6, no. 5, pp. B30-B36, May 1. 2018.

[73] X. Xue, Y. Xuan, C. Bao, S. Li, X. Zheng, B. Zhou, M. Qi, and A. M. Weiner, "Microcomb-Based True-Time-Delay Network for Microwave Beamforming With Arbitrary Beam Pattern Control," *Journal of Lightwave Technology,* vol. 36, no. 12, pp. 2312-2321, Jun. 2018.

[74] X. Xu, J. Wu, T. G. Nguyen, M. Shoeiby, S. T. Chu, B. E. Little, R. Morandotti, A. Mitchell, and D. J. Moss, "Advanced RF and microwave functions based on an integrated optical frequency comb source," *Optics Express,* vol. 26, no. 3, pp. 2569-2583, Feb 5. 2018.

[75] X. Xu, M. Tan, J. Wu, T. G. Nguyen, S. T. Chu, B. E. Little, R. Morandotti, A. Mitchell, and D. J. Moss, "Advanced Adaptive Photonic RF Filters with 80 Taps based on an integrated optical micro-comb," *Journal of Lightwave Technology,* vol. 37, no. 4, pp. 1288-1295, 2019.

[76] X. X. Xue, Y. Xuan, H. J. Kim, J. Wang, D. E. Leaird, M. H. Qi, and A. M. Weiner, "Programmable Single-Bandpass Photonic RF Filter Based on Kerr Comb from a Microring," *Journal of Lightwave Technology,* vol. 32, no. 20, pp. 3557-3565, Oct 15. 2014.

[77] X. Xu, J. Wu, M. Shoeiby, T. G. Nguyen, S. T. Chu, B. E. Little, R. Morandotti, A. Mitchell, and D. J. Moss, "Reconfigurable broadband microwave photonic intensity differentiator based on an integrated optical frequency comb source," *APL Photonics,* vol. 2, no. 9, 096104. Sept. 2017.

[78] M. Tan, X. Xu, B. Corcoran, J. Wu, A. Boes, T. G. Nguyen, S.T. Chu, B.E. Little, R. Morandotti, A. Mitchell, and D.J. Moss, "Microwave and RF photonic fractional Hilbert transformer based on a 50 GHz Kerr microcomb," *J. of Lightwave Technology,* vol. 37, no. 24, p 6097, 2019.

[79] M. Tan, X. Xu, B. Corcoran, J. Wu, A. Boes, T. G. Nguyen, S. T. Chu, B. E. Little, R. Morandotti, A. Mitchell, and D. J. Moss, "RF and microwave fractional differentiator based on photonics," IEEE Transactions on Circuits and Systems II: Express Briefs, Early Access. 2020. DOI: 10.1109/TCSII.2020.2965158

[80] X. Xu, J. Wu, M. Tan, T. G. Nguyen, S. Chu, B. Little, R. Morandotti, A. Mitchell, and D. J. Moss, "Micro-comb based photonic local oscillator for broadband microwave frequency conversion," *Journal of Lightwave Technology*, vol. 38, no. 2, pp. 332-338. 2020.

[81] X. Xu, M. Tan, J. Wu, A. Boes, B. Corcoran, T. G. Nguyen, S. T. Chu, B. Little, R. Morandotti, A. Mitchell and D. J. Moss, "Photonic RF phase-encoded signal generation with a microcomb source," *Journal of Lightwave Technology*, vol. 38, no.7, pp1722-1727. 2020.
DOI: 10.1109/JLT.2019.2958564

[82] X. Xu, M. Tan, J. Wu, T. G. Nguyen, S. T. Chu, B. E. Little, R. Morandotti, A. Mitchell, and D. J. Moss, "High performance RF filters via bandwidth scaling with Kerr micro-combs," *APL Photonics,* vol. 4, no. 2, pp. 026102. 2019.

[83] X. Xu, J. Wu, T. G. Nguyen, S. Chu, B. Little, A. Mitchell, R. Morandotti, and D. J. Moss, "Broadband RF Channelizer based on an Integrated Optical Frequency Kerr Comb Source," *Journal of Lightwave Technology,* vol. 36, no. 19, pp. 4519-4526. 2018.

[84] X.Xu, J.Wu, T.G. Nguyen, S.T. Chu, B.E. Little, R.Morandotti, A.Mitchell, and D. J. Moss, "Continuously tunable orthogonally polarized optical RF single sideband generator and equalizer based on an integrated microring resonator", IOP Journal of Optics, vol. 20 no.11, p115701, 2018.

[85] X.Xu, J.Wu, T.G.Nguyen, S.T.Chu, B.E.Little, R.Morandotti, A.Mitchell, and D.J. Moss, "Orthogonally polarized optical RF single sideband generator and equalizer based on an integrated micro-ring resonator", IEEE Journal of Lightwave Technology Vol. 36, No. 20, pp4808-4818 (2018).

[86] X.Xu, J.Wu, T.G. Nguyen, S.T. Chu, B.E. Little, R.Morandotti, A.Mitchell, and D. J. Moss, "Continuously tunable orthogonally polarized optical RF single sideband generator and equalizer based on an integrated microring resonator", IOP Journal of Optics, vol. 20 no.11, p115701, 2018.

[87] X.Xu, J.Wu, T.G.Nguyen, S.T.Chu, B.E.Little, R.Morandotti, A.Mitchell, and D.J. Moss, "Orthogonally polarized optical RF single sideband generator and equalizer based on an integrated micro-ring resonator", IEEE Journal of Lightwave Technology Vol. 36, No. 20, pp4808-4818 (2018).

[88] T.Ido, H.Sano, D.J.Moss, S.Tanaka, and A.Takai, "Strained InGaAs/InAlAs MQW electroabsorption modulators with large bandwidth and low driving voltage", Photonics Technology Letters, Vol. 6, 1207 (1994). DOI: 10.1109/68.329640.

[89] D. Moss, "Temporal RF photonic signal processing with Kerr micro-combs: Hilbert transforms, integration and fractional differentiation", OSF Preprints, 18 Feb. (2021). DOI: 10.31219/osf.io/hx9gb.

[90] D. Moss, "RF and microwave photonic high bandwidth signal processing based on Kerr micro-comb sources", TechRxiv. Preprint (2020). DOI:10.36227/techrxiv.12665609.v3.

[91] D. Moss, "RF and microwave photonic signal processing with Kerr micro-combs", Research Square (2021). DOI: 10.21203/rs.3.rs-473364/v1.

[92] M.Tan, X. Xu, J. Wu, D.J. Moss, "RF Photonic Signal Processing with Kerr Micro-Combs: Integration, Fractional Differentiation and Hilbert Transforms", Preprints (2020). 2020090597. doi:10.20944/preprints202009.0597.v1.

[93] Mengxi Tan, Xingyuan Xu, Jiayang Wu, Thach G. Nguyen, Sai T. Chu, Brent E. Little, Roberto Morandotti, Arnan Mitchell, and David J. Moss, "Photonic Radio Frequency Channelizers based on Kerr Micro-combs and Integrated Micro-ring Resonators", JOSarXiv.202010.0002.

[94] Mengxi Tan, Xingyuan Xu, David Moss "Tunable Broadband RF Photonic Fractional Hilbert Transformer Based on a Soliton Crystal Microcomb", Preprints, DOI: 10.20944/preprints202104.0162.v1

[95] Mengxi Tan, X. Xu, J. Wu, T. G. Nguyen, S. T. Chu, B. E. Little, R. Morandotti, A. Mitchell, and David J. Moss, "Orthogonally polarized Photonic Radio Frequency single sideband generation with integrated micro-ring resonators", Journal of Semiconductors vol. 42, No.4, 041305 (2021). DOI: 10.1088/1674-4926/42/4/041305.

[96] Mengxi Tan, X. Xu, J. Wu, T. G. Nguyen, S. T. Chu, B. E. Little, R. Morandotti, A. Mitchell, and David J. Moss, "Photonic Radio Frequency Channelizers based on Kerr Optical Micro-combs", Journal of Semiconductors 42 (4), 041302 (2021). (ISSN 1674-4926). DOI:10.1088/1674-4926/42/4/041302.

[97] Mengxi Tan, Xingyuan Xu, David Moss "Tunable Broadband RF Photonic Fractional Hilbert Transformer Based on a Soliton Crystal Microcomb", Preprints, DOI: 10.20944/preprints202104.0162.v1

[98] B. Corcoran, et al., "Ultra-dense optical data transmission over standard fiber with a single chip source", Nature Communications, vol. 11, Article:2568, 2020. DOI:10.1038/s41467-020-16265-x.

[99] X. Xu, et al., "Photonic perceptron based on a Kerr microcomb for scalable high speed optical neural networks", Laser and Photonics Reviews, vol. 14, no. 8, 2000070, 2020. DOI:10.1002/lpor.202000070.

[100] X. Xu, et al., "11 TOPs photonic convolutional accelerator for optical neural networks", Nature, vol.589 (7840) 44-51 (2021). DOI: 10.1038/s41586-020-03063-0.

[101] X Xu et al., "11 TeraFLOPs per second photonic convolutional accelerator for deep learning optical neural networks", arXiv preprint arXiv:2011.07393 (2020).

[102] D. Moss, "11 Tera-FLOP/s photonic convolutional accelerator and deep learning optical neural networks", Research Square (2021). DOI: https://doi.org/10.21203/rs.3.rs-493347/v1.

[103] D.Moss, "11.0 Tera-FLOP/second photonic convolutional accelerator for deep learning optical neural networks", TechRxiv. Preprint (2020). https://doi.org/10.36227/techrxiv.13238423.v1.

[104] Moss, David. "11 Tera-flop/s Photonic Convolutional Accelerator for Optical Neural Networks." OSF Preprints, 23 Feb. (2021). DOI: 10.31219/osf.io/vqt4s.

[105] Mengxi Tan, X. Xu, J. Wu, T. G. Nguyen, S. T. Chu, B. E. Little, R. Morandotti, A. Mitchell, and David J. Moss, "Photonic Radio Frequency Channelizers based on Kerr Optical Micro-combs", Journal of Semiconductors, Vol. 42, No. 4, 041302 (2021). (ISSN 1674-4926). DOI:10.1088/1674-4926/42/4/041302.

[106] H. Bao, L.Olivieri, M.Rowley, S.T. Chu, B.E. Little, R.Morandotti, D.J. Moss, J.S.T. Gongora, M.Peccianti and A. Pasquazi, "Laser Cavity Solitons and Turing Patterns in Microresonator Filtered Lasers: properties and perspectives", Paper No. LA203-5, Paper No. 11672-5, SPIE LASE, SPIE Photonics West, San Francisco CA March 6-11 (2021). DOI:10.1117/12.2576645

[107] Mengxi Tan, X. Xu, J. Wu, A. Boes, T. G. Nguyen, S. T. Chu, B. E. Little, R. Morandotti, A. Mitchell, and David J. Moss, "Advanced microwave signal generation and processing with soliton crystal microcombs", or "Photonic convolutional accelerator and neural network in the Tera-OPs regime based on Kerr microcombs", Paper No. 11689-38, PW21O-OE201-67, Integrated Optics: Devices, Materials, and Technologies XXV, SPIE Photonics West, San Francisco CA March 6-11 (2021). DOI: 10.1117/12.2584017

[108] Mengxi Tan, Bill Corcoran, Xingyuan Xu, Andrew Boes, Jiayang Wu, Thach Nguyen, Sai T. Chu, Brent E. Little, Roberto Morandotti, Arnan Mitchell, and David J. Moss, "Optical data transmission at 40 Terabits/s with a Kerr soliton crystal microcomb", Paper No.11713-8, PW21O-OE803-23, Next-Generation Optical Communication: Components, Sub-Systems, and Systems X, SPIE Photonics West, San Francisco CA March 6-11 (2021). DOI:10.1117/12.2584014

[109] Mengxi Tan, X. Xu, J. Wu, A. Boes, T. G. Nguyen, S. T. Chu, B. E. Little, R. Morandotti, A. Mitchell, and David J. Moss, "RF and microwave photonic, fractional differentiation, integration, and Hilbert transforms based on Kerr micro-combs", Paper No. 11713-16, PW21O-OE803-24, Next-Generation Optical Communication: Components, Sub-Systems, and Systems X, SPIE Photonics West, San Francisco CA March 6-11 (2021). DOI:10.1117/12.2584018

[110] Mengxi Tan, X. Xu, J. Wu, A. Boes, T. G. Nguyen, S. T. Chu, B. E. Little, R. Morandotti, A. Mitchell, and David J. Moss, "Broadband photonic RF channelizer with 90 channels based on a soliton crystal microcomb", or "Photonic microwave and RF channelizers based on Kerr micro-combs", Paper No. 11685-22, PW21O-OE106-49, Terahertz, RF, Millimeter, and Submillimeter-Wave Technology and Applications XIV, SPIE Photonics West, San Francisco CA March 6-11 (2021). DOI:10.1117/12.2584015

[111] X. Xu, M. Tan, J. Wu, S. T. Chu, B. E. Little, R. Morandotti, A. Mitchell, B. Corcoran, D. Hicks, and D. J. Moss, "Photonic perceptron based on a Kerr microcomb for scalable high speed optical neural networks", IEEE Topical Meeting on Microwave Photonics (MPW), pp. 220-224, Matsue, Japan, November 24-26, 2020. Electronic ISBN:978-4-88552-331-1. DOI: 10.23919/MWP48676.2020.9314409

[112] Mengxi Tan, Bill Corcoran, Xingyuan Xu, Andrew Boes, Jiayang Wu, Thach Nguyen, S.T. Chu, B. E. Little, Roberto Morandotti, Arnan Mitchell, and David J. Moss, "Ultra-high bandwidth optical data transmission with a microcomb", IEEE Topical Meeting on Microwave Photonics (MPW), pp. 78-82. Virtual Conf., Matsue, Japan, November 24-26, 2020. Electronic ISBN:978-4-88552-331-1. DOI: 10.23919/MWP48676.2020.9314476

[113] M. Tan, X. Xu, J. Wu, R. Morandotti, A. Mitchell, and D. J. Moss, "RF and microwave high bandwidth signal processing based on Kerr Micro-combs", Advances in Physics X, VOL. 6, NO. 1, 1838946 (2020). DOI:10.1080/23746149.2020.1838946.

[114] M Tan, X Xu, J Wu, DJ Moss, "High bandwidth temporal RF photonic signal processing with Kerr micro-combs: integration, fractional differentiation and Hilbert transforms", arXiv preprint arXiv:2103.03674 (2021).

[115] Mengxi Tan, X. Xu, J. Wu, T. G. Nguyen, S. T. Chu, B. E. Little, R. Morandotti, A. Mitchell, and David J. Moss, "Orthogonally polarized Photonic Radio Frequency single sideband generation with integrated micro-ring resonators", Journal of Semiconductors 42 (4), 041305 (2021). DOI: 10.1088/1674-4926/42/4/041305.

[116] L. Razzari, D. Duchesne, M. Ferrera, et al., "CMOS-compatible integrated optical hyper-parametric oscillator," Nature Photonics, vol. 4, no. 1, pp. 41-45 (2010).
[117] M. Ferrera, L. Razzari, D. Duchesne, et al., "Low-power continuous-wave nonlinear optics in doped silica glass integrated waveguide structures," Nature Photonics, vol. 2, no. 12, pp. 737-740 (2008).
[118] Pasquazi, et al., "Sub-picosecond phase-sensitive optical pulse characterization on a chip", Nature Photonics, vol. 5, no. 10, pp. 618-623 (2011). DOI: 10.1038/nphoton.2011.199.
[119] D. Duchesne, M. Peccianti, M. R. E. Lamont, et al., "Supercontinuum generation in a high index doped silica glass spiral waveguide," Optics Express, vol. 18, no, 2, pp. 923-930 (2010).
[120] M. Ferrera, et al., "On-chip CMOS-compatible all-optical integrator", Nature Communications, vol. 1, Article 29 (2010).
[121] H. Bao et al., "Turing patterns in a fibre laser with a nested micro-resonator: robust and controllable micro-comb generation", Physical Review Research, vol. 2, pp. 023395 (2020).
[122] L. D. Lauro, J. Li, D. J. Moss, R. Morandotti, S. T. Chu, M. Peccianti, and A. Pasquazi, "Parametric control of thermal self-pulsation in micro-cavities," Opt. Lett. vol. 42, no. 17, pp. 3407-3410, Aug. 2017.
[123] H. Bao et al., "Type-II micro-comb generation in a filter-driven four wave mixing laser," Photonics Research, vol. 6, no. 5, pp. B67-B73 (2018).
[124] Pasquazi, et al., "All-optical wavelength conversion in an integrated ring resonator," Optics Express, vol. 18, no. 4, pp. 3858-3863 (2010).
[125] Pasquazi, Y. Park, J. Azana, et al., "Efficient wavelength conversion and net parametric gain via Four Wave Mixing in a high index doped silica waveguide," Optics Express, vol. 18, no. 8, pp. 7634-7641 (2010).
[126] M. Peccianti, M. Ferrera, L. Razzari, et al., "Subpicosecond optical pulse compression via an integrated nonlinear chirper," Optics Express, vol. 18, no. 8, pp. 7625-7633 (2010).
[127] D. Duchesne, M. Ferrera, L. Razzari, et al., "Efficient self-phase modulation in low loss, high index doped silica glass integrated waveguides," Optics Express, vol. 17, no. 3, pp. 1865-1870 (2009).
[128] M. Peccianti, et al., "Demonstration of an ultrafast nonlinear microcavity modelocked laser", Nature Communications, vol. 3, pp. 765 (2012).
[129] M. Kues, et al., "Passively modelocked laser with an ultra-narrow spectral width", Nature Photonics, vol. 11, no. 3, pp. 159 (2017). DOI:10.1038/nphoton.2016.271
[130] Pasquazi, L. Caspani, M. Peccianti, et al., "Self-locked optical parametric oscillation in a CMOS compatible microring resonator: a route to robust optical frequency comb generation on a chip," Optics Express, vol. 21, no. 11, pp. 13333-13341 (2013).
[131] Pasquazi, M. Peccianti, B. E. Little, et al., "Stable, dual mode, high repetition rate mode-locked laser based on a microring resonator," Optics Express, vol. 20, no. 24, pp. 27355-27362 (2012).
[132] Reimer, L. Caspani, M. Clerici, et al., "Integrated frequency comb source of heralded single photons," Optics Express, vol. 22, no. 6, pp. 6535-6546 (2014).
[133] Reimer, et al., "Cross-polarized photon-pair generation and bi-chromatically pumped optical parametric oscillation on a chip", Nature Communications, vol. 6, Article 8236 (2015). DOI: 10.1038/ncomms9236
[134] L. Caspani, et al., "Multifrequency sources of quantum correlated photon pairs on-chip: a path toward integrated Quantum Frequency Combs," Nanophotonics, vol. 5, no. 2, pp. 351-362 (2016).
[135] Reimer, M. Kues, P. Roztocki, B. Wetzel, F. Grazioso, B. E. Little, S. T. Chu, T. Johnston, Y. Bromberg, L. Caspani, D. J. Moss, and R. Morandotti, "Generation of multiphoton entangled quantum states by means of integrated frequency combs," Science, vol. 351, no. 6278, pp. 1176-1180 (2016).
[136] P. Roztocki, M. Kues, C. Reimer, B. Wetzel, S. Sciara, Y. Zhang, A. Cino, B. E. Little, S. T. Chu, D. J. Moss, and R. Morandotti, "Practical system for the generation of pulsed quantum frequency combs," Optics Express, vol. 25, no. 16, pp. 18940-18949 (2017).
[137] Y. Zhang, et al., "Induced photon correlations through superposition of two four-wave mixing processes in integrated cavities", Laser and Photonics Reviews, vol. 14, no. 7, 2000128 (2020). DOI: 10.1002/lpor.202000128
[138] M. Kues, C. Reimer, A. Weiner, J. Lukens, W. Munro, D. J. Moss, and R. Morandotti, "Quantum Optical Micro-combs", Nature Photonics, vol. 13, no.3, pp. 170-179 (2019).
[139] Reimer, et al.,"High-dimensional one-way quantum processing implemented on d-level cluster states", Nature Physics, vol. 15, no.2, pp. 148–153 (2019).
[140] H. Arianfard et al., "Three waveguide coupled sagnac loop reflectors for advanced spectral engineering," J. Lightwave Technol., doi: 10.1109/JLT.2021.3066256.

[141] J. Wu et al., "Nested configuration of silicon microring resonator with multiple coupling regimes," IEEE Photon. Technol. Lett., vol. 25, no. 6, pp. 580-583, Mar. 2013.
[142] J. Wu, T. Moein, X. Xu, and D. J. Moss, "Advanced photonic filters based on cascaded Sagnac loop reflector resonators in silicon-on-insulator nanowires," APL Photonics, vol. 3, 046102 (2018). DOI:/10.1063/1.5025833Apr. 2018.
[143] J Wu, T Moein, X Xu, DJ Moss, "Silicon photonic filters based on cascaded Sagnac loop resonators", arXiv:1805.05405a (2018).
[144] J. Wu, T. Moein, X. Xu, G. H. Ren, A. Mitchell, and D. J. Moss, "Micro-ring resonator quality factor enhancement via an integrated Fabry-Perot cavity," APL Photonics, vol. 2, 056103 (2017).
[145] H. Arianfard, J. Wu, S. Juodkazis, and D. J. Moss, "Advanced Multi-Functional Integrated Photonic Filters Based on Coupled Sagnac Loop Reflectors", Journal of Lightwave Technology, Vol. 39, No.5, pp.1400-1408 (2021). DOI: 10.1109/JLT.2020.3037559.
[146] David J. Moss, "Optimization of Optical Filters based on Integrated Coupled Sagnac Loop Reflectors", Research Square (2021). DOI: https://doi.org/10.21203/rs.3.rs-478204/v1
[147] H. Arianfard, J. Wu, S. Juodkazis, D. J. Moss, "Spectral Shaping Based on Integrated Coupled Sagnac Loop Reflectors Formed by a Self-Coupled Wire Waveguide", submitted, IEEE Photonics Technology Letters, vol. 33 (2021).
[148] J. Wu et al., "Graphene oxide waveguide and micro-ring resonator polarizers," Laser Photonics Rev., vol. 13, no. 9, pp. 1900056, Aug. 2019.
[149] Y. Zhang et al., "Optimizing the Kerr nonlinear optical performance of silicon waveguides integrated with 2D graphene oxide films," J. Lightwave Technol., doi: 10.1109/JLT.2021.3069733.
[150] Y. Qu et al., "Analysis of four-wave mixing in silicon nitride waveguides integrated with 2D layered graphene oxide films," J. Lightwave Technol., vol. 39, no. 9, pp. 2902-2910, May. 2021.
[151] Y. Qu et al., "Enhanced four-wave mixing in silicon nitride waveguides integrated with 2D layered graphene oxide films," Adv. Opt. Mater., vol. 8, no. 20, pp. 2001048, Oct. 2020.
[152] J. Wu et al., "2D layered graphene oxide films integrated with micro-ring resonators for enhanced nonlinear optics," Small, vol. 16, no. 16, pp. 1906563, Mar. 2020.
[153] Moss, David; Wu, Jiayang; xu, xingyuan; Yang, Yunyi; jia, linnan; Zhang, Yuning; et al. (2020): Enhanced optical four-wave-mixing in integrated ring resonators with graphene oxide films. TechRxiv. Preprint. https://doi.org/10.36227/techrxiv.11859429.v1.
[154] Wu, J.; Yang, Y.; Qu, Y.; Jia, L.; Zhang, Y.; Xu, X.; Chu, S.T.; Little, B.E.; Morandotti, R.; Jia, B.; Moss, D.J. Enhancing Third Order Nonlinear Optics in Integrated Ring Resonators with 2D Material Films. Preprints 2020, 2020030107
[155] Wu, J.; Yang, Y.; Qu, Y.; Jia, L.; Zhang, Y.; Xu, X.; Chu, S.T.; Little, B.E.; Morandotti, R.; Jia, B.; Moss, D.J. Enhancing Third Order Nonlinear Optics in Integrated Ring Resonators with 2D Material Films. Preprints 2020, 2020030107
[156] J Wu et al., "Enhanced four-wave-mixing with 2D layered graphene oxide films integrated with CMOS compatible micro-ring resonators", arXiv preprint, arXiv:2002.04158 (2020).
[157] Y. Zhang et al., "Enhanced Kerr nonlinearity and nonlinear figure of merit in silicon nanowires integrated with 2D graphene oxide films," ACS Appl. Mater. Interfaces, vol. 12, no. 29, pp. 33094-33103, Jun. (2020).
[158] TD Vo et al., "Silicon-chip-based real-time dispersion monitoring for 640 Gbit/s DPSK signals", Journal of Lightwave Technology, vol. 29, no. 12, 1790-1796 (2011).
[159] C Grillet, C Smith, D Freeman, S Madden, B Luther-Davies, EC Magi, "Efficient coupling to chalcogenide glass photonic crystal waveguides via silica optical fiber nanowires", Optics Express, vol. 14, no. 3, pp.1070-1078 (2006).
[160] Y Zhang et al., "Enhanced nonlinear optical figure-of-merit at 1550nm for silicon nanowires integrated with graphene oxide layered films", arXiv preprint arXiv:2004.08043 (2020).
[161] Moss et al., "Transforming silicon into a high performing integrated nonlinear photonics platform by integration with 2D graphene oxide films", TechRxiv. Preprint. (2020). https://doi.org/10.36227/techrxiv.12061809.v1.
[162] Moss, "Elevating silicon into a high performance nonlinear optical platform through the integration of 2D graphene oxide thin films", Research Square (2021). DOI: https://doi.org/10.21203/rs.3.rs-511259/v1.
[163] Moss, D.; Wu, J.; Jia, B.; Yang, Y.; Qu, Y.; Jia, L.; Zhang, Y. "Improved Nonlinear Optics in Silicon-on-insulator Nanowires Integrated with 2D Graphene Oxide Films", Preprints (2020), 2020040033.
[164] Y. Yang et al., "Invited article: enhanced four-wave mixing in waveguides integrated with graphene oxide," APL Photonics, vol. 3, no. 12, pp. 120803, Oct. 2018.